\documentclass[11pt,a4paper]{amsart}
\usepackage{amsmath}
\usepackage{amssymb}
\usepackage{amsfonts}
\usepackage[toc,page]{appendix}
\usepackage{hyperref}
\usepackage{color}

\def\={\ =\ }

\newcommand{\be}{\begin{equation}}
\newcommand{\ee}{\end{equation}}
\newcommand{\beq}{\begin{equation}}
\newcommand{\eeq}{\end{equation}}
\newcommand{\bea}{\begin{eqnarray}}
\newcommand{\eea}{\end{eqnarray}}

\def\ba{\begin{eqnarray}}
\def\ea{\end{eqnarray}}

\theoremstyle{plain}

\numberwithin{equation}{section}

\newcommand{\func}[1]{\operatorname{#1}}

\input{tcilatex}

\begin{document}
\title[]{Mass-deformed ABJ and ABJM theory, Meixner-Pollaczek polynomials,
and $su(1,1)$ oscillators}
\author{}
\author{Miguel Tierz}
\address{Departamento de Matem\'{a}tica, Grupo de F\'{\i}sica Matem\'{a}%
tica, Faculdade de Ci\^{e}ncias, Universidade de Lisboa, Campo Grande, Edif%
\'{\i}cio C6, 1749-016 Lisboa, Portugal.}
\email{tierz@fc.ul.pt}
\maketitle

\begin{abstract}
We give explicit analytical expressions for the partition function of $%
U(N)_{k}\times U(N+M)_{-k}$ ABJ theory at weak coupling ($k\rightarrow
\infty )$ for finite and arbitrary values of $N$ and $M$ (including the ABJM
case and its mass-deformed generalization). We obtain the expressions by
identifying the one-matrix model formulation with a Meixner-Pollaczek
ensemble and using the corresponding orthogonal polynomials, which are also
eigenfunctions of a $su(1,1)$ quantum oscillator. Wilson loops in
mass-deformed ABJM are also studied in the same limit and interpreted in
terms of $su(1,1)$ coherent states.
\end{abstract}

\section{Introduction}

In recent years there has been considerable progress in applications of the
localization technique to the study of supersymmetric gauge theories in a
number of dimensions, $d\geq 2$. In three dimensions in particular, results
have been obtained for the partition functions and BPS Wilson loops of $%
\mathcal{N}=2$ supersymmetric Chern-Simons-matter (CSM) theories, including
the $\mathcal{N}=6$ superconformal theories constructed by Aharony, Bergman,
Jafferis and Maldacena, (ABJM theory) \cite{Aharony:2008ug}. We shall focus
here in the 3d realm exclusively and, in particular and more specifically,
on the partition function of the ABJ theory \cite{Aharony:2008gk}, which is
the $\mathcal{N}=6$ supersymmetric $U(N_{1})_{k}\times U(N_{2})_{-k}$ CSM
theory that generalizes the equal rank $N_{1}=N_{2}$ case of the ABJM theory.

In addition to the extension of ABJM theory, it has also been conjectured
that the ABJ theory at large $N_{2}$ and $k$ with $N_{2}/k$ and $N_{1}$
fixed finite is dual to the $\mathcal{N}=6$ parity-violating Vasiliev higher
spin theory on $AdS_{4}$ with $U(N_{1})$ gauge symmetry \cite%
{Chang:2012kt,Hirano:2015yha,Honda:2015sxa}. In this paper, we will be studying the ABJ
theory at large $k$ and hence we shall be commenting on this higher-spin
scaling limit.

The application of the localization method results in the reduction of path
integrals into eigenvalue integrals of the matrix model type, which allows
to obtain various exact results at strong coupling of supersymmetric gauge
theories and helps in providing further tests of the AdS/CFT correspondence.
The appearance of matrix models typically implies a connection with exactly
solvable models and/or integrable systems. This will be the viewpoint of
this paper, where we identify ABJ theory at large $k$ with an exactly
solvable model. More precisely with a Meixner-Pollaczek random matrix
ensemble \cite{Kan}, solved completely in terms of Meixner-Pollaczek
orthogonal polynomials \cite{orth-book}, which are also eigenfunctions of
the $su(1,1)$ harmonic oscillator \cite{Jafarov:2012qq}.

Using either the two-matrix model description of ABJ(M) theory or the Fermi
gas formulation, both briefly described below, a large number of results
have been obtained, mostly in studying non-perturbative corrections of the
ABJ(M) matrix models. Both the\ 't Hooft expansion, which is the asymptotic
expansion as $N$ goes to infinity and the 't Hooft parameter of the model is
kept fixed at large $N$, and the M-theory expansion, which is the asymptotic
expansion as $N$ goes to infinity in which $k$ is fixed, have now been
studied in detail, see \cite%
{Marino:2011nm,Grassi:2014vwa,Codesido:2014oua,Hirano:2015yha,Honda:2015sxa} for example.

The ABJ(M) $U(N)_{k}\times U(N+M)_{-k}$ two-matrix model is given by \cite%
{Marino:2011nm,Awata:2012jb}%
\begin{eqnarray}
Z_{\mathrm{ABJ(M)}}(N,k,M) &=&\mathcal{N}_{\mathrm{ABJ}}\iint \frac{{d}^{N}{%
\mu }}{\left( 2\pi \right) ^{N}}\frac{d^{N+M}y}{\left( 2\pi \right) ^{N+M}}%
\frac{\prod\limits_{i<j}4\sinh ^{2}\left( \frac{1}{2}\left( \mu _{i}-\mu
_{j}\right) \right) \prod\limits_{i<j}4\sinh ^{2}\left( \frac{1}{2}\left(
y_{i}-y_{j}\right) \right) }{\prod_{i=1}^{N}\prod_{j=1}^{N+M}\left( 2\cosh
\left( \frac{1}{2}(\mu _{i}-y_{j})\right) \right) ^{2}}  \notag \\
&&\times e^{\frac{ik}{4\pi }\left( \sum_{i=1}^{N}\mu
_{i}^{2}-\sum_{i=1}^{N+M}y_{i}^{2}\right) },  \label{ZABJM1}
\end{eqnarray}%
where%
\begin{equation*}
\mathcal{N}_{\mathrm{ABJ}}=\frac{i^{-\kappa M(N-1/2)}}{N!(N+M)!}\text{ \ \ \
\ \ \ where }\kappa :=sgn(k).
\end{equation*}%
The ABJM case corresponds to $M=0$ above. By identifying the Cauchy
determinant inside the integrand of (\ref{ZABJM1}) and using Cauchy identity 
\begin{equation}
\frac{\prod\limits_{i<j}\sinh ^{2}\left( \frac{1}{2}\left( \mu _{i}-\mu
_{j}\right) \right) \prod\limits_{i<j}\sinh ^{2}\left( \frac{1}{2}\left(
y_{i}-y_{j}\right) \right) }{\prod_{i=1}^{N}\prod_{j=1}^{N+M}\left( \cosh
\left( \frac{1}{2}(\mu _{i}-y_{j})\right) \right) ^{2}}=\sum_{\sigma \in
S_{N}}\left( -1\right) ^{\varepsilon \left( \sigma \right) }\prod\limits_{i}%
\frac{1}{\cosh \left( \mu _{i}-y_{\sigma (i)}\right) },  \label{cauchydet}
\end{equation}%
one can write the matrix model as a sum of permutations \cite%
{Kapustin:2010xq,Marino:2011eh}. In turn, in the final expression obtained,
which is the basis of the Fermi gas method \cite{Marino:2011eh}, one can
apply again the Cauchy determinant identity, obtaining then the one-matrix
model form of the ABJ(M) matrix model \cite%
{Awata:2012jb,Honda:2013pea,Codesido:2014oua}%
\begin{equation}
\widetilde{Z}_{\mathrm{ABJ(M)}}(N,k,M)=\frac{1}{N!}\int \prod_{j=1}^{N}\frac{%
dx_{j}}{4\pi k}\frac{\prod_{l=-(M-1)/2}^{(M-1)/2}\tanh \frac{x_{j}+2\pi il}{k%
}}{e^{x_{j}/2}+(-1)^{M}e^{-x_{j}/2}}\prod_{i<j}\tanh ^{2}(\frac{1}{2k}%
(x_{i}-x_{j})).  \label{tanhmodel}
\end{equation}%
Note that the matrix model integral has $N$ eigenvalues and the dependence
in $M$ is through the potential. This is the matrix model we shall study,
focussing on the large $k$ limit. In the ABJM case, $M=0$ above, the matrix
integral (\ref{tanhmodel}) gives the full partition function but in the more
general ABJ case a somewhat lengthy numerical prefactor, which includes the
partition function of $U(M)$ Chern-Simons theory on $S^{3}$ appears as well 
\cite[(2.14)]{Honda:2013pea}. We shall focus on the matrix integral (\ref%
{tanhmodel}) first and will discuss these additional prefactors at the end
of the next Section.

This paper is organized as follows. In the next Section, we show that the
large $k$ limit of the one matrix model formulation of ABJ theory is a
Meixner-Pollaczek random matrix ensemble and compute the partition functions
for finite $N$ and $M$ (from now on, we always write $N_{2}=N+M$). Some
quotients between partition functions, relevant in the study of the
higher-spin limit, are given and shown to coincide with the partition
function of the Penner matrix model.

In Section 3, we show that the identification with a solvable model also
holds for the mass-deformed version of the theories, focussing on the
mass-deformed ABJM theory.\ An analytical computation for this model is also
given, using the orthogonal polynomials, which admit an interpretation as
eigenfunctions of a $su(1,1)$ quantum oscillator. In this interpretation,
the $M$ parameter is equal to the Bargmann index of the positive discrete
series representation of $su(1,1)$.\ We also study 1/6-BPS Wilson loops in
the fundamental representation with winding $n$, in the mass-deformed ABJM
theory, giving an interpretation in terms of $su(1,1)$ coherent states. The
discussion of Wilson loops in Section 3 is carried out in comparison with $%
\frac{1}{2}$-BPS Wilson loop in $\mathcal{N}=4$ super Yang-Mills theory,
whose coherent state interpretation is also put forward.

Finally, we succinctly conclude with some avenues for further research. In
the Appendix we collect some technical details on the analytical
continuations in the parameters (Section 2 and 3) and on the analytical
solution for Wilson loops (Section 3).

\section{Weak coupling limit and Meixner-Pollaczek polynomials}

Let us start focussing on (\ref{tanhmodel}) in the ABJM case, $M=0.$ Then,
the weight function of the matrix model (\ref{tanhmodel}), which in general
is%
\begin{equation*}
\omega (x;M)=e^{-V(x;M)}=\frac{\prod_{l=-(M-1)/2}^{(M-1)/2}\tanh \frac{%
x+2\pi il}{k}}{e^{x/2}+(-1)^{M}e^{-x/2}},
\end{equation*}%
greatly simplifies to%
\begin{equation}
\omega (x;M=0)=\frac{1}{2\cosh \left( x/2\right) }.  \label{abjmweight}
\end{equation}%
The most distinctive mathematical feature of the matrix model (\ref%
{tanhmodel}) is the interaction term between the eigenvalues. It admits,
like the model with a standard Vandermonde term $\prod_{i<j}(x_{i}-x_{j})$
or with its hyperbolic version $\prod_{i<j}\sinh (x_{i}-x_{j})$, a Coulomb
gas interpretation \cite{tanh}, but in general there are no standardized
analytical methods to solve the resulting matrix model. Consequently, the
one matrix model formulation has been analyzed with less detail than the
original two-matrix model formulation (\ref{ZABJM1}) or the Fermi gas
formulation \cite{Marino:2011eh}.

A simple yet fundamental regime which has only eventually been analyzed is
the weak-coupling limit $k\rightarrow \infty $, when $N$ is fixed.
Expressions for the partition function in the literature include, for the
ABJM case, given in \cite{Marino:2011nm} and a computation in \cite[Appendix
C]{Awata:2012jb} for ABJ theory, using (\ref{ZABJM1}) and the analytical
continuation from a lens space Chern-Simons matrix model.

To study this regime, the consideration of the one-matrix model formulation (%
\ref{tanhmodel}) is especially useful. Indeed, the very simple dependence of 
$k$ in the matrix model (\ref{tanhmodel}), immediately implies that, using
the rescaled variable $\pi y=x/2$%
\begin{equation}
Z_{\mathrm{ABJM}}(N,k\rightarrow \infty )=\frac{1}{N!}\frac{\pi ^{N(N-1)}}{%
2^{2N}k^{N^{2}}}\int \prod_{i=1}^{N}\frac{dy_{i}}{\cosh \left( \pi
y_{i}\right) }\prod_{i<j}(y_{i}-y_{j})^{2},  \label{semabjm}
\end{equation}%
which is in the form of a standard random matrix ensemble. In addition, the
weight function of the model (\ref{semabjm}) can also be easily identified
with a particular case of the Meixner-Pollaczek polynomials \cite%
{Kan,orth-book}, which are polynomials orthogonal in the interval $-\infty
<x<\infty $, with regards to the weight function \cite{Kan}\footnote{%
This is actually not the most general weight function for this system of
polynomials, since there is also a phase factor, but we will only need that
case below, when we discuss the mass-deformed ABJM model.}%
\begin{equation}
w^{(\lambda )}(x)=\frac{2^{2\lambda -1}}{\pi }\left\vert \Gamma (\lambda
+ix)\right\vert ^{2}=\frac{2^{2\lambda -1}\left\vert \Gamma \left( \lambda
\right) \right\vert ^{2}}{\pi }\prod_{k=0}^{\infty }\left( 1+\frac{x^{2}}{%
\left( k+\lambda \right) ^{2}}\right) ^{-1}.  \label{A2}
\end{equation}%
Notice that the Gamma function is evaluated in a complex argument and that%
\begin{equation*}
\left\vert \Gamma (\lambda +ix)\right\vert ^{2}=\Gamma (\lambda +ix)\Gamma
(\lambda -ix)\text{ \ \ \ for \ \ }x,\lambda \in 
\mathbb{R}
.
\end{equation*}%
Therefore, for $\lambda =1/2,$ and also taking into account Euler's
duplication formula $\Gamma (z)\Gamma (1-z)=\pi /\sin \left( \pi z\right) $,
then%
\begin{equation*}
w^{(1/2)}(x)=\frac{1}{\cosh (\pi x)}=\omega _{\mathrm{ABJM}}(x),
\end{equation*}%
and with the orthogonal polynomials one can compute immediately (\ref%
{semabjm}). We do that below, but before we discuss the more general ABJ
case (\ref{tanhmodel}) because, interestingly enough, this identification
with an exactly solvable model also holds with more generality, as it holds
for the ABJ matrix model. In particular, we have that, again in terms of the 
$y$ variable%
\begin{equation}
\widetilde{Z}_{\mathrm{ABJ}}(N,k\rightarrow \infty ,M)=\frac{1}{N!}\frac{%
2^{N(M-1)}\pi ^{N^{2}-N+MN}}{k^{N(N+M)}}\int \prod_{j=1}^{N}\frac{%
\prod_{l=-(M-1)/2}^{(M-1)/2}\left( y_{j}+il\right) dy_{j}}{e^{\pi
y_{j}}+(-1)^{M}e^{-\pi y_{j}}}\prod_{i<j}(y_{i}-y_{j})^{2}.  \label{ABJMM}
\end{equation}%
To simplify the presentation, we focus first on the case when $M=2q$ and $%
q\in 
\mathbb{N}
$ and later on we show that the odd case $M=2q-1$ works in the same way.
Then, the weight function of the matrix model is%
\begin{equation*}
\omega _{\mathrm{ABJ}}^{\mathrm{(even)}}(y)=\frac{\prod_{l=1/2}^{(M-1)/2}%
\left( y^{2}+l^{2}\right) }{2\cosh (\pi y)}.
\end{equation*}%
This is precisely the Meixner-Pollaczek weight if we take its parameter $%
\lambda $ to be $\lambda =1/2+M/2=1/2+q$ with $q\in 
\mathbb{N}
$. This follows immediately from the form (\ref{A2}) and $\Gamma
(z+1)=z\Gamma (z)$. Specifically:%
\begin{eqnarray}
\widetilde{Z}_{\mathrm{ABJ}}(N,k &\rightarrow &\infty ,M)=\frac{1}{N!}\frac{%
2^{N(M-2)}\pi ^{N^{2}-N+MN}}{k^{N(N+M)}}\int \prod_{i=1}^{N}\frac{%
\prod_{l=1/2}^{(M-1)/2}\left( y_{i}^{2}+l^{2}\right) dy_{i}}{\cosh (\pi
y_{i})}\prod_{i<j}(y_{i}-y_{j})^{2}  \notag \\
&=&\frac{1}{N!}\frac{\pi ^{N^{2}-N+MN}}{2^{2N}k^{N(N+M)}}\int
\prod_{i=1}^{N}w^{(1/2+M/2)}(y_{i})dy_{i}\prod_{i<j}(y_{i}-y_{j})^{2},
\end{eqnarray}%
where the weight function is (\ref{A2}) with $\lambda =1/2+M/2$.

Let us take into account then the orthogonality properties of the
polynomials, which is all we need to obtain the partition functions/free
energies. The polynomials $P_{n}^{(\lambda )}(x)$ satisfy \cite{Kan}%
\begin{equation}
\int_{-\infty }^{+\infty }dxP_{n}^{(\lambda )}(x)P_{m}^{(\lambda
)}(x)w^{(\lambda )}(x)=\delta _{nm}h_{n}^{(\lambda )},  \label{A3}
\end{equation}%
with%
\begin{equation}
h_{n}^{(\lambda )}=\frac{\Gamma (n+2\lambda )}{\Gamma (n+1)}.  \label{A4}
\end{equation}%
From the recurrence relation \cite{orth-book}, the leading coefficient of
the polynomial $P_{n}^{(\lambda )}(x)=a_{n}x^{n}+...$ is also obtained%
\begin{equation}
a_{n}=\frac{2^{n}}{\Gamma \left( n+1\right) }.  \label{A5}
\end{equation}%
Then, we can use, exactly as with the Stieltjes-Wigert polynomials and the
computation of the free energy of $U(N)$\ Chern-Simons theory on $S^{3}$ 
\cite{Tierz:2002jj}, the explicit analytical expression for the partition
function of the matrix model, using%
\begin{equation*}
Z=N!\prod\limits_{j=0}^{N-1}\widetilde{h}_{j},
\end{equation*}%
where $Z$ denotes the partition function of a Hermitian matrix model with
the Meixner-Pollaczek weight function and $\widetilde{h}_{j}$ denote the
analogous of the $h_{j}$ in (\ref{A3}), but for the \emph{monic} orthogonal
polynomials. Thus, we need the orthogonality properties of the polynomials $%
Q_{n}^{(\lambda )}(x)=P_{n}^{(\lambda )}(x)/a_{n}$ , which are%
\begin{equation}
\int_{-\infty }^{+\infty }dxQ_{n}^{(\lambda )}(x)Q_{m}^{(\lambda
)}(x)w^{(\lambda )}(x)=\delta _{nm}\widetilde{h}_{n}^{(\lambda )}=\delta
_{nm}\frac{h_{n}^{(\lambda )}}{a_{n}^{2}}.  \label{A3b}
\end{equation}%
Therefore, we obtain for (\ref{semabjm})%
\begin{eqnarray*}
Z_{\mathrm{ABJM}}(N,k &\rightarrow &\infty )=\frac{\pi ^{N^{2}-N}}{%
2^{2N}k^{N^{2}}}\prod\limits_{j=0}^{N-1}\frac{h_{j}^{(1/2)}}{a_{j}^{2}}=%
\frac{\pi ^{N^{2}-N}}{2^{2N}k^{N^{2}}}\prod\limits_{j=0}^{N-1}4^{-j}\Gamma
^{2}\left( j+1\right)  \\
&=&\frac{\pi ^{N^{2}-N}}{2^{N\left( N+1\right) }k^{N^{2}}}G^{2}(N+1)=\left( 
\frac{\pi }{2k}\right) ^{N^{2}}\frac{G^{2}(N+1)}{\left( 2\pi \right) ^{N}},
\end{eqnarray*}%
where $G(z)$ is a Barnes G-function \cite{G}. Hence, the free energy $F=\ln Z
$ is%
\begin{equation}
F_{\mathrm{ABJM}}(N,k\rightarrow \infty )=N^{2}\log \left( \frac{\pi }{2k}%
\right) -N\log (2\pi )+2\log G(N+1).  \label{Fabjm}
\end{equation}%
This expression is identical to the one given in \cite[Eq. 4.47]%
{Marino:2011nm}, which is obtained there, since it is a perturbative
computation, by evaluation of the determinants giving the one-loop
contributions of two copies of pure Chern-Simons theory and the one loop
determinants of the matter fields. The free energy is the sum of both
contributions and this is why the expression is organized in a slightly
different way there. Thus, the Meixner-Pollaczek polynomials capture exactly
both contributions together. For the ABJ partition function, we obtain%
\begin{equation}
\widetilde{Z}_{\mathrm{ABJ}}(N,M,k\rightarrow \infty )=\frac{\pi
^{N^{2}-N+MN}}{2^{2N}k^{N(N+M)}}\prod\limits_{j=0}^{N-1}\frac{%
h_{j}^{(1/2+M/2)}}{a_{j}^{2}}=\frac{\pi ^{N^{2}-N+MN}}{2^{N(N+1)}k^{N(N+M)}}%
\frac{G(N+M+1)G(N+1)}{G(M+1)}.  \label{ABJresult1a}
\end{equation}%
Notice that, in this weak coupling limit, it holds that%
\begin{equation}
\frac{\widetilde{Z}_{\mathrm{ABJ}}(N,M,k\rightarrow \infty )}{Z_{\mathrm{ABJM%
}}(N,k\rightarrow \infty )}=\left( \frac{\pi }{k}\right) ^{^{NM}}\frac{%
G(N+M+1)}{G(N+1)G(M+1)}.  \label{quotient}
\end{equation}%
Below we will show how this quotient is modified when we consider the full
ABJ partition function. Since $\widetilde{Z}_{\mathrm{ABJ}}$ does not
include the pure Chern-Simons partition function factor, the ratio (\ref%
{quotient}) is precisely the one considered to be specially relevant in the
higher-spin double scaling limit \cite[Eq. 5.2.]{Hirano:2015yha} (dubbed
\textquotedblleft vector model subsector\textquotedblright\ in \cite%
{Hirano:2015yha}). The r.h.s. of (\ref{quotient}) is also, exactly, the
partition function of the Penner matrix model (a Laguerre random matrix
ensemble), which is characterized by a weight function $\omega (x)=x^{M}\exp
(-x)$ for $x>0$. It would be interesting to see if this can be related both
to the results in \cite[Eq. 5.2.]{Hirano:2015yha} and, at the same time,
with the well-known results on the $c=1$ string, topological strings, which
are associated with the Penner matrix model (see \cite{Mukhi:2003sz} for a
review).

We consider now the case of $M$ odd since we restricted the discussion above
to $M$ being an even number just to simplify the presentation of the
connection with Meixner-Pollaczek polynomials. It also holds for $M$ odd in
the same way. The difference being that, instead of half-integer values of $%
\lambda $, we will have integer values. This is immediate from, again,
Euler's duplication formula, because%
\begin{equation*}
\left\vert \Gamma (1+ix)\right\vert ^{2}=\Gamma (1+ix)\Gamma (1-ix)=\frac{%
\pi x}{\sinh \pi x},
\end{equation*}%
therefore, increasing $\lambda $ by $1$ adds a $(1+x^{2})$ term, and so on.
The correspondence with the semiclassical limit of the ABJ weight is then
again exact%
\begin{equation*}
\omega _{\mathrm{ABJ}}^{\mathrm{(odd)}}(y)=\frac{\prod_{l=0}^{(M-1)/2}\left(
y^{2}+l^{2}\right) }{\sinh (\pi y)}=2^{-\left( M-1\right)
}w^{(1+(M-1)/2)}(y),
\end{equation*}%
and, therefore, for the $M$ odd case, we have%
\begin{equation*}
\widetilde{Z}_{\mathrm{ABJ}}(N,k\rightarrow \infty ,M)=\frac{\pi
^{N^{2}-N+MN}}{2^{2N}k^{N(N+M)}}\prod\limits_{j=0}^{N-1}\frac{%
h_{j}^{(1+\left( M-1\right) /2)}}{a_{j}^{2}}=\frac{\pi ^{N^{2}-N+MN}}{%
2^{2N}k^{N(N+M)}}\prod\limits_{j=0}^{N-1}\frac{\Gamma (j+1+M)\Gamma
^{2}\left( j+1\right) }{4^{j}},
\end{equation*}%
which is the same result as for $M$ even and, therefore (\ref{ABJresult1a})
holds for $M$ any natural number and the identification is always $\lambda
=1/2+M/2$ for $M\in 
\mathbb{N}
$. Below we will see that this parameter can be interpreted as the Bargmann
index of the positive discrete series representation of the $su(1,1)$ Lie
algebra.

Let us now finally deal with the extra factors that multiply (\ref{tanhmodel}%
). The exact relationship between (\ref{ZABJM1}) and (\ref{tanhmodel}) is
detailed in \cite{Honda:2013pea}, with $q=\exp (2i\pi /k)$%
\begin{eqnarray}
&&Z_{\mathrm{ABJ}}^{(N,N+M)}(k)=\frac{i^{-\frac{\mathrm{sign}(k)}{2}%
(N^{2}+(N+M)^{2})}(-1)^{\frac{N}{2}(N-1)+\frac{M}{2}(M-1)+NM}i^{N+\frac{M}{2}%
}q^{\frac{M}{12}(M^{2}-1)}}{k^{\frac{M}{2}}}  \notag \\
&&\times \prod_{1\leq l<m\leq M}2i\sin {\frac{\pi (l-m)}{k}}
\label{ABJ_mirror} \\
&&\times \frac{1}{N!}\int_{-\infty }^{\infty }\frac{d^{N}y}{(4\pi k)^{N}}%
\prod_{a<b}\tanh ^{2}{\frac{y_{a}-y_{b}}{2k}}\prod_{a=1}^{N}\frac{1}{2\cosh {%
\frac{y_{a}}{2}}}\prod_{l=0}^{M-1}\tanh {\frac{y_{a}+2\pi i(l+1/2)}{2k}}. 
\notag
\end{eqnarray}%
We can write, for simplicity, the $k$-independent prefactor in the first
line in (\ref{ABJ_mirror}) as $\overline{\mathcal{N}}_{\mathrm{ABJ}}(N,M)$
and the most interesting aspect of considering the full prefactor in (\ref%
{ABJ_mirror}) is that the $k\rightarrow \infty $ of what is essentially the $%
U(M)$ Chern-Simons partition function on $S^{3}$ gives a $G(M+1)$ Barnes
function which cancels the one in (\ref{ABJresult1a}). More precisely%
\begin{eqnarray*}
\lim_{k\rightarrow \infty }Z_{\mathrm{ABJ}}^{(N,N+M)}(k) &=&k^{\frac{M(M-2)}{%
2}}\overline{\mathcal{N}}_{\mathrm{ABJ}}(N,M)G(M+1)\widetilde{Z}_{\mathrm{ABJ%
}}(N,M,k\rightarrow \infty ) \\
&=&\frac{\overline{\mathcal{N}}_{\mathrm{ABJ}}(N,M)\pi ^{N^{2}-N+MN}k^{\frac{%
M(M-2)}{2}}}{2^{N(N+1)}k^{N(N+M)}}G(N+M+1)G(N+1),
\end{eqnarray*}%
where%
\begin{equation*}
\overline{\mathcal{N}}_{\mathrm{ABJ}}(N,M)=i^{-\frac{\mathrm{sign}(k)}{2}%
(N^{2}+(N+M)^{2})}(-1)^{\frac{N}{2}(N-1)+\frac{M}{2}(M-1)+NM}i^{N+\frac{M^{2}%
}{2}}\left( 2\pi \right) ^{\frac{M}{2}(M-1)}.
\end{equation*}%
The quotient (\ref{quotient}) between the full ABJ and ABJM partition
function is now 
\begin{equation}
\frac{Z_{\mathrm{ABJ}}(N,M,k\rightarrow \infty )}{Z_{\mathrm{ABJM}%
}(N,k\rightarrow \infty )}=\overline{\mathcal{N}}_{\mathrm{ABJ}}k^{\frac{%
M(M-2)}{2}}\left( \frac{\pi }{k}\right) ^{^{NM}}\frac{G(N+M+1)}{G(N+1)}.
\label{realquotient}
\end{equation}%
In addition to the usual asymptotics of the $G$ Barnes function, one can
give an asymptotic expression for the quotient of $G$ functions in (\ref%
{realquotient}) for $N$ and $M$ finite, but $N$ much bigger than $M$. This
estimates, in compact form, how the ABJ free energy differs from the ABJM
one at small values of $M$. More precisely, from \cite[Prop 17. (i)]{Kow},
we have that, for $M\leq N^{1/6}$ then%
\begin{equation}
\frac{G(1+M+N)}{G(1+N)}=(2\pi )^{M/2}e^{-(N+1)M}(1+N)^{M^{2}/2+NM}\Bigl(1+%
\mathcal{O}\Bigl(\frac{M^{2}+M^{3}}{N}\Bigr)\Bigr).  \label{eq-power-ratio}
\end{equation}%
Thus, we have seen that with Meixner-Pollaczek polynomials, which are
actually eigenfunctions of the $su(1,1)$ harmonic oscillator as we shall see
below, the finite $N$ large $k$ behavior of the ABJ free energy is easily
obtained. This complements the varied number of results in the literature on
ABJ(M) models and the expressions obtained are consistent with \cite%
{Marino:2011nm} and \cite{Awata:2012jb}, giving also an alternative
derivation for the ABJ case, to the one in \cite{Awata:2012jb}, without
having to rely on analytical continuation from the lens space matrix model.
We can also apply the method to a mass-deformed version of the theory and
that is the subject of the next Section.

\section{Mass-deformed theory,\ Wilson loops and $su(1,1)$ coherent states}

More relevant than the application itself is the fact that, starting with
the one matrix model formulation (\ref{tanhmodel}), the ABJ matrix model,
for all $M$ and $N$, is an exactly solvable model in the weak-coupling
limit: the Meixner-Pollaczek model. Further results can be obtained from
that identification. For example, it can be extended to the mass-deformed
ABJM matrix model, which has been the subject of interest lately. It was
first studied at large $N$ in \cite{Anderson:2014hxa} (with Fayet-Iliopoulos
parameter $\varsigma =0$). The model is \cite{AR,Russo:2015exa} 
\begin{equation}
Z(N,k,m_{1},m_{2})=\frac{1}{N!^{2}}\iint \frac{{d}^{N}{\mu }}{\left( 2\pi
\right) ^{N}}\frac{d^{N}y}{\left( 2\pi \right) ^{N}}\frac{%
\prod\limits_{i<j}\sinh ^{2}\left( \frac{\mu _{i}-\mu _{j}}{2}\right)
\prod\limits_{i<j}\sinh ^{2}\left( \frac{y_{i}-y_{j}}{2}\right) e^{\frac{ik}{%
4\pi }\sum_{i=1}^{N}\left( \mu _{i}^{2}-y_{i}^{2}\right) }}{%
\prod_{i,j=1}^{N}\cosh \left( \frac{1}{2}(\mu _{i}-y_{j}+m_{1})\right) \cosh
\left( \frac{1}{2}(\mu _{i}-y_{j}-m_{2})\right) }.  \label{mdef}
\end{equation}%
This extension of the ABJM matrix model actually also includes a
Fayet-Iliopoulos parameter $\varsigma $, which has been repackaged with the
mass term deformation as $m_{1}=m+\varsigma $ and $m_{2}=m-\varsigma $ \cite%
{AR,Russo:2015exa}.

It also admits a sum over permutations form \cite{AR,Russo:2015exa} and, if
one again applies to the resulting expression the Cauchy determinant
formula, we see that we also have a one matrix model form of the
mass-deformed ABJM matrix model\footnote{%
A factor of 4 is added in the denominator of the matrix model w.r.t the
expressions in \cite{AR,Russo:2015exa}. In this way, in the limit $%
m_{1}\rightarrow 0$ and $m_{2}\rightarrow 0$ it reduces to the ABJM matrix
model above, which is as in \cite{Grassi:2014vwa}, for example. The reason
is that the two $\cosh $ in \cite{Grassi:2014vwa}, unlike those in \cite%
{AR,Russo:2015exa}, have a 2 term in front.}%
\begin{eqnarray}
Z_{\mathrm{ABJM}}(N,k,m_{1},m_{2}) &=&\frac{1}{N!}\int \prod_{i=1}^{N}\frac{%
e^{-im_{2}x_{i}}dx_{i}}{4k\cosh \left( \pi x_{i}\right) }\frac{%
\prod_{i<j}\sinh ^{2}(\frac{\pi }{k}(x_{i}-x_{j}))}{\prod_{i,j}\cosh (\frac{%
\pi }{k}(x_{i}-x_{j})-\frac{m_{1}}{2})}  \notag \\
&=&\frac{1}{N!\cosh \left( m_{1}/2\right) ^{N}2^{2N}k^{N}}\int
\prod_{i=1}^{N}\frac{e^{-im_{2}x_{i}}dx_{i}}{\cosh \left( \pi x_{i}\right) }%
\prod_{i<j}\frac{\sinh ^{2}(\frac{\pi }{k}(x_{i}-x_{j}))}{\cosh ^{2}(\frac{%
\pi }{k}(x_{i}-x_{j})-\frac{m_{1}}{2})}.
\end{eqnarray}%
Thus, in first approximation in the semiclassical limit, we have%
\begin{equation}
Z_{\mathrm{ABJM}}(N,k\rightarrow \infty ,m_{1},m_{2})=\frac{\pi ^{N^{2}-N}}{%
N!\cosh \left( m_{1}/2\right) ^{N^{2}}4^{N}k^{N^{2}}}\int \prod_{i=1}^{N}%
\frac{e^{-im_{2}x_{i}}dx_{i}}{\cosh \left( \pi x_{i}\right) }%
\prod_{i<j}(x_{i}-x_{j})^{2}.  \label{mABJM}
\end{equation}%
Notice that, to account for the $m_{1}$ deformation is immediate in this way
and, if we also want to account for the $m_{2}$ parameter, we need to
consider what is actually the most general form of the Meixner-Pollaczek
weight function \cite{orth-book}%
\begin{equation}
w_{\mathrm{MP}}(x;\lambda ,t)=\frac{2^{2\lambda -1}e^{tx}}{\pi }\left\vert
\Gamma (\lambda +ix)\right\vert ^{2},  \label{MPweight}
\end{equation}%
where $t\in \left( -\pi ,\pi \right) $. This restriction on $t$ is because,
for fixed $\lambda $, $\left\vert \Gamma (\lambda +ix)\right\vert ^{2}\sim
e^{-\pi \left\vert x\right\vert }$ as $\left\vert x\right\vert \rightarrow
\infty $, therefore it ensures exponential convergence for $x\rightarrow \pm
\infty $. Being $t$ real we need to take its analytical prolongation to
imaginary values at the end (or, alternatively, do the same to the physical
parameter $m_{2}$). Such an step has precedents in the application of
orthogonal polynomial methods to the study of Chern-Simons theory, with or
without matter \cite{Tierz:2002jj,Tierz:2016zcn}\footnote{%
The works \cite{Amariti:2011hw,Amariti:2011da} also consider orthogonal
polynomials in Chern-Simons-matter theory.} and, as we shall see, it
poses no problems here. In the case $m_{2}=0,$ this continuation is not
necessary and the parameter $\phi $ below will be $\phi =$ $\pi /2$, as is
the case for the non-deformed ABJM theory. As explained in \cite%
{Russo:2015exa}, this specific mass-deformation is a fixed point of the
symmetry $Z(2\varsigma ,m;k)=Z(m,2\varsigma ;k)$, satisfied by the
mass-deformed theory and, in the dual $\mathcal{N}$ $=4$ supersymmetric
super Yang-Mills theory, $m_{2}=0$ corresponds to coupling the theory to a
massless adjoint hypermultiplet.

In most references, the notation convention is $t=2\phi -\pi $, where $\phi
\in \left( 0,\pi \right) $ and the orthogonality properties now read%
\begin{equation}
\frac{2^{2\lambda -1}}{\pi }\int_{-\infty }^{+\infty }e^{\left( 2\phi -\pi
\right) x}\left\vert \Gamma (\lambda +ix)\right\vert ^{2}P_{n}^{(\lambda
)}(x;\phi )P_{m}^{(\lambda )}(x;\phi )dx=\frac{\Gamma (n+2\lambda )}{\left(
\sin \phi \right) ^{2\lambda }\Gamma (n+1)}\delta _{nm}.  \label{nA3}
\end{equation}%
Notice that, if $\phi =\pi /2,$ the weight function indeed reduces to (\ref%
{A2}). We need to see if the leading coefficients of the polynomial, which
did not depend on $\lambda $, depend now on $\phi $. From the recurrence
relationships \cite{orth-book}, we quickly deduce that%
\begin{equation}
a_{n}^{\left( \phi \right) }=\frac{\left( 2\sin \phi \right) ^{n}}{\Gamma
\left( n+1\right) }.  \label{kphi}
\end{equation}%
Thus, since the weight function of the mass-deformed ABJM model (\ref{mABJM}%
) is $w_{\mathrm{MP}}(x;\lambda =1/2,t=2\phi -\pi =-im_{2})$, then using (%
\ref{nA3}) and (\ref{kphi}), we have, in terms of the $\phi $ parameter%
\begin{eqnarray*}
Z_{\mathrm{ABJM}}(N,m_{1},\phi ,k &\rightarrow &\infty )=\left( \frac{\pi }{%
k\cosh \left( m_{1}/2\right) }\right) ^{N^{2}}\frac{1}{\left( 4\pi \right)
^{N}}\prod\limits_{j=0}^{N-1}\frac{h_{j,\phi }^{(1/2)}}{a_{j,\phi }^{2}} \\
&=&\frac{1}{\left( 2\pi \right) ^{N}}\left( \frac{\pi }{2k\cosh \left(
m_{1}/2\right) }\right) ^{N^{2}}\frac{G^{2}(N+1)}{\left( \sin \phi \right)
^{N^{2}}}.
\end{eqnarray*}%
The identification $\phi =-im_{2}/2+\pi /2$ poses no problems, as we shall
explicitly check in the Appendix, and we have%
\begin{equation*}
Z_{\mathrm{ABJM}}(N,m_{1},m_{2},k\rightarrow \infty )=\left( \frac{\pi }{%
2k\cosh \left( m_{1}/2\right) \cosh \left( m_{2}/2\right) }\right) ^{N^{2}}%
\frac{G^{2}(N+1)}{\left( 2\pi \right) ^{N}}.
\end{equation*}%
Thus, in this limit, for the mass deformed model we have that the free
energy 
\begin{equation*}
F_{\mathrm{ABJM}}(m_{1},m_{2})=F_{\mathrm{ABJM}}-N^{2}(\ln \cosh \left(
m_{1}/2\right) +\ln \cosh \left( m_{2}/2\right) ).
\end{equation*}

\subsection{Wilson loops in the mass-deformed case}

We consider Wilson loops now. The idea is to see if the Wilson loop can be
computed as an average using a density of states constructed from quantum
oscillators, very much as the Drukker-Gross computation of the $\frac{1}{2}$%
-BPS Wilson loop in $\mathcal{N}=4$ super Yang-Mills theory \cite%
{Drukker:2000rr}, which uses the ordinary quantum harmonic (Hermite)
wavefunctions. To do so, we will use the Meixner-Pollaczek polynomials and
their $su(1,1)$ quantum oscillator interpretation, giving also a coherent
state interpretation, both for the mass-deformed ABJ Wilson loops at weak
coupling and also for the $\frac{1}{2}$-BPS Wilson loop in $\mathcal{N}=4$
super Yang-Mills theory \cite{Drukker:2000rr}.

For this, one first needs to write down the one matrix model representation
of the Wilson loop. We have not seen that result in the literature, at least
not in a form convenient to our purpose and methods (there are essentially
equivalent integral representations for Wilson loops in \cite%
{Honda:2013pea,Hirano:2014bia}). Thus, we work out the required one matrix
model expression, by repeating the procedure in \cite[Eq. 5-Eq. 8]%
{Anderson:2014hxa} for the ABJM mass deformed theory, but with a Wilson loop
insertion. The Wilson loop averages we consider are the $\frac{1}{6}$-BPS
Wilson loops in \cite{Kapustin:2009kz}%
\begin{equation}
\left\langle W_{R}^{1/6}\right\rangle =\left\langle \mathrm{Tr}_{R}(e^{\mu
})\right\rangle ,  \label{ABJMW}
\end{equation}%
where the average is taken over the ensemble (\ref{ZABJM1}) with $M=0$ and
one could equivalently consider the Wilson loop over the other gauge group.
The $\frac{1}{2}$-BPS Wilson loops of Drukker and Trancanelli treat both
gauge groups more symmetrically \cite{Drukker:2009hy}, but they can be
expressed in terms of the $\frac{1}{6}$-BPS Wilson loops. We will consider (%
\ref{ABJMW}) in the fundamental representation and with winding $n$%
\begin{equation}
\left\langle W_{n}^{1/6}\right\rangle _{m}=\left\langle \mathrm{Tr}(e^{n\mu
})\right\rangle _{m},  \label{nWm}
\end{equation}%
and the ensemble average will be over the mass-deformed ABJM model (\ref%
{mdef}), instead of the regular ABJM. This is denoted by the $m$ subindex in
the average.

The procedure in \cite[Eq. 5-Eq. 8]{Anderson:2014hxa} consists in using the
Fourier transform identity%
\begin{equation}
\int d\tau \frac{e^{i\tau \mu }}{\cosh \left( \pi \tau \right) }=\frac{1}{%
\cosh \frac{\mu }{2}},  \label{fu}
\end{equation}%
for (\ref{nWm}) twice (as an integral representation; that is, from right to
left), a Gaussian integration and a further immediate integration, using (%
\ref{fu}). We do exactly the same for (\ref{nWm}), which is explicitly given
by the matrix model (\ref{mdef}) with a $\sum_{j=1}^{N}e^{n\mu _{j}}$
insertion in the integrand. We obtain that the one-matrix model
representation of the Wilson loop is%
\begin{equation}
\left\langle W_{n}^{1/6}\right\rangle _{m}=\frac{c_{n}}{Z}\int
\prod_{i=1}^{N}\frac{e^{-im_{2}x_{i}}\sum_{j=1}^{N}e^{\frac{2\pi n}{k}x_{j}}%
}{4k\cosh \left( \pi x_{i}\right) }dx_{i}\frac{\prod_{i<j}\sinh ^{2}(\frac{%
\pi }{k}(x_{i}-x_{j}))}{\prod_{i,j}\cosh (\frac{\pi }{k}(x_{i}-x_{j})-\frac{%
m_{1}}{2}+\frac{i\pi n}{k}\delta _{ij})},  \label{W-exp}
\end{equation}%
where the prefactor $c_{n}$ is generated by the Wilson loop insertion and is
given by $c_{n}=e^{-in^{2}\pi /k-n\varsigma }$. Notice the appearance of $k$
in the added exponential insertion in the integrand in (\ref{W-exp}) (this
is due to the Gaussian integration in the procedure). In the $k\rightarrow
\infty $ we can then take the double-scaling limit with the winding $n$ such
that $n/k=\beta $, and therefore, we have that%
\begin{equation}
\lim_{\substack{ k\rightarrow \infty  \\ n/k\rightarrow \beta }}\left\langle
W_{n}^{1/6}\right\rangle _{m}=\frac{\alpha _{N}(k,\beta ,m_{1})c_{n}}{Z}\int
\prod_{i=1}^{N}\frac{e^{-im_{2}x_{i}}\sum_{j=1}^{N}e^{2\pi \beta x_{j}}}{%
\cosh \left( \pi x_{i}\right) }dx_{i}\prod_{i<j}(x_{i}-x_{j})^{2},
\label{Wmav}
\end{equation}%
where the prefactor $\alpha _{N}(k,\beta ,m_{1})$ follows in the same way as
for the partition function (\ref{mABJM}). Specifically:%
\begin{equation*}
\alpha _{N}(k,\beta ,m_{1})=\frac{\pi ^{N^{2}-N}}{\cosh \left( m_{1}/2-i\pi
\beta \right) ^{N}\cosh \left( m_{1}/2\right) ^{N(N-1)}4^{N}k^{N^{2}}}.
\end{equation*}%
We first given an analytical expression for the the matrix integral in (\ref%
{Wmav}) and, in the next Subsection, a physical interpretation. Notice that,
in general, for a Hermitian matrix model with weight function $\omega \left(
x\right) $%
\begin{equation}
\frac{\left\langle \mathrm{Tr}e^{yM}\right\rangle }{Z}=\sum_{s=0}^{N-1}\frac{%
\int dx\omega \left( x\right) e^{yx}P_{s}^{2}(x)}{\int dx\omega \left(
x\right) P_{s}^{2}(x)}=\sum_{s=0}^{N-1}\left\langle s\left\vert
e^{yx}\right\vert s\right\rangle ,  \label{avH}
\end{equation}%
where%
\begin{equation}
\left\langle x\mid s\right\rangle =\frac{\omega \left( x\right)
^{1/2}P_{s}(x)}{\left[ \int dx\omega \left( x\right) P_{s}^{2}(x)\right]
^{1/2}}  \label{pols}
\end{equation}%
and $P_{s}(x)$ denotes the polynomial of order $s$, orthogonal w.r.t. $%
\omega \left( x\right) $. Thus, it is as in the $\mathcal{N}=4$ SYM/Hermite
setting \cite{Drukker:2000rr} but with an imaginary $y$ parameter, instead
of a real one. Hence, denoting by $\left\vert n;\lambda ,\phi \right\rangle $
the Meixner-Pollaczek eigenstates\footnote{%
These are the states such that the r.h.s. of (\ref{pols}) is the normalized
Meixner-Pollaczek polynomial of degree $s$ and the weight function is (\ref%
{MPweight}).}, we have that the non-trivial piece of the Wilson loop (that
is, up to normalization constants) is%
\begin{equation}
\lim_{\substack{ k\rightarrow \infty  \\ n/k\rightarrow \mu }}\left\langle
W_{n}^{1/6}\right\rangle _{m}\propto \sum\limits_{n=0}^{N-1}\left\langle
n;\lambda =1/2,\phi \left\vert \exp (-im_{2}\widehat{X})\right\vert
n;\lambda =1/2,\phi \right\rangle ,  \label{WSum}
\end{equation}%
with $\phi =i\pi (\frac{1}{2}-\beta )$ and $\widehat{X}$ is the position
operator. Below, we will identify the operator insertion in (\ref{WSum})
with a squeeze operator, but we focus first on obtaining an analytical
expression. The explicit evaluation of (\ref{WSum}) follows from a Fourier
integral in \cite{MCS}, which generalizes the orthogonality identity of the
MP polynomials (\ref{nA3}) and is the analogue of the identity for Hermite
polynomials which leads to (\ref{matrix}) \cite{Drukker:2000rr} (see
Appendix for further details). In particular, using \cite{MCS}, we have the
following expression in terms of a (terminating) Gauss hypergeometric
function%
\begin{eqnarray}
&&\left\langle n;\lambda =1/2,\phi \left\vert \exp (-im_{2}\widehat{X}%
)\right\vert n;\lambda =1/2,\phi \right\rangle   \label{W} \\
&=&_{2}F_{1}\left( 
\begin{array}{c}
-n,-n \\ 
1%
\end{array}%
;-\left( \frac{\cos \left( \pi \beta \right) }{\sinh \left( \frac{m_{2}}{2}%
\right) }\right) ^{2}\right) \frac{\pi i\left( \sinh \left( \frac{m_{2}}{2}%
\right) \right) ^{2n}}{\left( \sinh \left( \frac{m_{2}}{2}+i\pi (\frac{1}{2}%
-\beta \right) \right) ^{2n+1}}.  \notag
\end{eqnarray}%
In this expression, we have identified the Fourier kernel in (\ref{Fourier})
with the mass-deformed term in the matrix integral $e^{-im_{2}x}$, and the
Wilson loop insertion with the real exponential part $e^{\left( 2\phi -\pi
\right) x}$ of the weight function. Hence, no analytical continuation is
necessary and one can also give an expression for the Wilson loop in the
standard ABJM theory with only the standard orthogonality relationship (\ref%
{nA3}). Notice that, in contrast to the Hermite case, the evaluation of the
summation in (\ref{WSum}) of the hypergeometric terms (\ref{W}), to obtain a
more compact expression for the Wilson loop, is an open problem. On the
other hand, specific cases at finite rank are immediate from (\ref{W}),
since the hypergeometric expression above is a terminating series, and a
more detailed study, including extension to the ABJ setting and comparison
with perturbative computations, will be given elsewhere. 

\subsection{$su(1,1)$ oscillator, coherent states and photonic interpretation%
}

We now exploit a physical interpretation of the Meixner-Pollaczek
polynomials: they are eigenfunctions of a quantum oscillator model. Indeed,
the $su(1,1)$ model of a quantum oscillator is a model which obeys the
dynamics of a harmonic oscillator, but with the position and momentum
operators and the Hamiltonian being elements of the Lie algebra $su(1,1)$
instead of the Heisenberg algebra \cite{Jafarov:2012qq}. For an oscillator
model, one requires also that the spectrum of $H$ in unitary representations
of the Lie algebra is equidistant. The generators of the algebra are $K_{\pm
}$ and $K_{0}$ and their commutation relations 
\begin{equation}
\left[ K_{-},K_{+}\right] =2K_{0}\text{ \ \ and \ \ }\left[ K_{0},K_{\pm }%
\right] =\pm K_{\pm }  \label{gen}
\end{equation}%
The model is constructed using the positive discrete series representations
of $su(1,1)$, $D^{+}(\lambda )$, which are infinite-dimensional and labeled
by a positive number (Bargmann index) $\lambda >0$. Then the spectrum of the
position operator is $%
\mathbb{R}
$, the spectra of the Hamiltonian is $n+\lambda $ and the position wave
function, when the oscillator is in the $n$th eigenstate of the Hamiltonian,
is given by \cite{Jafarov:2012qq} 
\begin{equation}
\phi _{n}\left( x;\lambda ,\phi \right) =C_{n}e^{\left( 2\phi -\pi \right)
x}\left\vert \Gamma (\lambda +ix)\right\vert ^{2}P_{n}^{(\lambda )}(x;\phi ),
\label{MPw}
\end{equation}%
where again $P_{n}^{(\lambda )}(x;\phi )$ are Meixner-Pollaczek polynomials.
Since the partition function of a Hermitian matrix model is%
\begin{equation*}
Z_{N}=\int \rho _{N}\left( x\right) dx=\int \lim_{x^{\prime }\rightarrow
x}K_{N}\left( x^{\prime },x\right) dx,
\end{equation*}%
where $\rho _{N}\left( x\right) $ is the density of states and the two-point
kernel is%
\begin{equation*}
K_{N}\left( x^{\prime },x\right) =\left( \omega \left( x^{\prime }\right)
\omega \left( x\right) \right) ^{1/2}\sum_{n=0}^{N-1}P_{n}^{(\lambda
)}(x^{\prime };\phi )P_{n}^{(\lambda )}(x;\phi ).
\end{equation*}%
Therefore, we have an immediate interpretation in terms of a $su(1,1)$
oscillator wavefunction overlap 
\begin{equation*}
\widetilde{Z}_{\mathrm{ABJ}}(N,M,k\rightarrow \infty )=\left\langle \Phi
\mid \Phi \right\rangle ,
\end{equation*}%
where $\left\vert \Phi \right\rangle =\sum_{n=0}^{N-1}\left\vert
n\right\rangle $ and 
\begin{equation*}
\left\langle x\mid n\right\rangle =\widetilde{\phi }_{n}\left( x,\lambda
=1/2+M/2,\phi =\pi /2\right) ,
\end{equation*}%
where $\widetilde{\phi }_{n}\left( x,\lambda =1/2+M/2,\phi =\pi /2\right) $
is (\ref{MPw}), properly normalized. That is, with $C_{n}=\parallel \phi
_{n}\parallel _{2}^{-1}$. Likewise, the same expression holds for the
mass-deformed ABJM case above, but with normalized eigenfunctions 
\begin{equation*}
\left\langle x\mid n\right\rangle =\widetilde{\phi }_{n}\left( x,\lambda
=1/2,\phi =-im_{2}/2+\pi /2\right) .
\end{equation*}%
From the results above, we thus have a half-integer Bargmann index for $M$
even and integer for $M$ odd. The case $\lambda =1/2$, which is the one that
corresponds to ABJM theory, including its mass-deformed version, is
precisely the one that has the exact quantum oscillator spectra $n+1/2$,
without any shift in the ground state energy.

This relationship with a $su(1,1)$ quantum oscillator can be further
developed in the case of Wilson loops. For this, notice first that, for a
Gaussian matrix model, the (\ref{pols}) are Hermite polynomials, and the
resulting expression for the Wilson loop average (\ref{avH}) is that of
Drukker and Gross \cite{Drukker:2000rr}. In terms of the creation and
annihilation operators $a=\lambda /2+d/d\lambda $ and $a^{\dagger }=$ $%
\lambda /2-d/d\lambda $ the average (\ref{avH}) is that of the displacement
operator \cite{Cahill:1969it} 
\begin{equation*}
D(y)\equiv \exp (y(a+a^{\dagger })).
\end{equation*}
The displacement operator acting on the vacuum state generates a coherent
state $D(y)\left\vert 0\right\rangle =\left\vert y\right\rangle $ and when
it acts on an excited state $D(y)\left\vert n\right\rangle $ these are
displaced number states \cite{Nieto:1996eh}, which are also wavepackets that
keep their shape and follow classical motion. The evaluation of the matrix
element in (\ref{avH}) is a central result in the theory of coherent states
and quantum optics \cite{Cahill:1969it} and in laser cooling \cite{Wineland}%
. Its evaluation is classical (and can be completely algebraic, without
relying on Hermite polynomials identities) \cite{Cahill:1969it,Wineland}%
\begin{equation}
\left\langle n^{\prime }\left\vert e^{y(a+a^{\dagger })}\right\vert
n\right\rangle =e^{-y^{2}/2}y^{\Delta n}\left( \sqrt{\frac{n_{<}!}{n_{>}!}}%
\right) L_{n_{<}}^{\Delta n}(-y^{2}),  \label{matrix}
\end{equation}%
where $n_{<}=\min (n,n^{\prime })$, $n_{>}=\max (n,n^{\prime })$ and $\Delta
n=n_{>}-n_{<}$. Comparing matrix models, the generic $y$ parameter in (\ref%
{avH}) here is identified with the t' Hooft parameter in \cite%
{Drukker:2000rr} $y=\sqrt{\lambda /4N}$ and then (\ref{matrix}) for $%
n=n^{\prime }$ coincides with the result in \cite{Drukker:2000rr}\footnote{%
In the work \cite{Okuyama:2006jc} on half-BPS Wilson loops in $\mathcal{N}=4$
SYM theory, the displacement operator is introduced and coherent states
mentioned, but mostly \textit{en route} to obtain a normal matrix model
description of the Wilson loops.}. Because of the summation property of
Laguerre polynomials $\sum\nolimits_{n=0}^{N-1}L_{n}^{\alpha
}(x)=L_{n}^{\alpha +1}(x)$ (used in \cite{Drukker:2000rr}), the Wilson loops
can also be expressed as a single matrix element (\ref{matrix}) between
adjacent levels $n^{\prime }=n+1=N$. That is 
\begin{equation}
\left\langle W_{\mathcal{N}=4}^{1/2}\right\rangle =2N\lambda
^{-1/2}\left\langle N\left\vert e^{\sqrt{\lambda /4N}(a+a^{\dagger
})}\right\vert N-1\right\rangle .  \label{newW-D}
\end{equation}%
This displacement operator average in turn can be interpreted as the average
of a unitary operator involving a Jaynes-Cummings Hamiltonian, but such a
quantum optics interpretation of the $\frac{1}{2}$-BPS Wilson loop in $%
\mathcal{N}=4$ super Yang-Mills theory (\ref{newW-D}) shall be better
discussed elsewhere.

Regarding our weak-coupling ABJ model, we have that the averages in (\ref%
{WSum}) can be interpreted as averages of the $su(1,1)$ displacement operator%
\footnote{%
The reason is simply the same that leads to (\ref{matrix}) in the quantum
harmonic oscillator case. Notice that the position operator is $\widehat{X}%
=(K_{+}+K_{-})/2$ and that these generators are raising and lowering
operators acting on the $su(1,1)$ oscillator states \cite{Jafarov:2012qq}.},
which is, in terms of the generators (\ref{gen}) and with $\xi \in 
\mathbb{C}
$ \cite{Nieto:1996eh} 
\begin{equation}
S(\xi )=\exp \left( \xi ^{\ast }K_{-}-\xi K_{+}\right) ,  \label{D11}
\end{equation}%
acting on the states of the $su(1,1)$ quantum oscillator and with the
identification $\xi =-im_{2}/2$, as explained above.

For a photonic interpretation and to further understand the meaning of the
displacement operator average in the $su(1,1)$ setting, recall that there is
a well-known two-mode realization\footnote{%
The one boson realization only allows for Bargmann index $\lambda =1/4$ or $%
\lambda =3/4$.} of $su(1,1)$ (the Holstein-Primakoff realization for $su(1,1)
$) where%
\begin{eqnarray*}
K_{+} &=&a^{\dagger }b^{\dagger }, \\
K_{-} &=&ab, \\
K_{0} &=&\frac{1}{2}\left( a^{\dagger }a+b^{\dagger }b+1\right) .
\end{eqnarray*}%
The states corresponding to the discrete positive series are then given by $%
\left\vert n+n_{0},n\right\rangle =\left\vert n+n_{0}\right\rangle \otimes
\left\vert n\right\rangle $, where the Bargmann index is $\lambda
=(\left\vert n_{0}\right\vert +1)/2$. Thus, the observables in the large $k$
limit of the ABJ theory can be described in terms of states of a two-mode
photonic system with occupancies of $n$ and $n+n_{0}$ photons in each mode
with $n=0,1,2...,N-1$ and $n_{0}=2M-1$ for $M=1,2,...$ and $n_{0}=0$ for $M=0
$ (ABJM). Notice also that the $su(1,1)$ displacement operator (\ref{D11})
is $e^{\xi ab-\xi ^{\ast }a^{\dagger }b^{\dagger }}$and hence it is a
squeeze operator acting on these two-mode states (generalized coherent
states for $su(1,1)$ are conventional squeezed states of quantum optics).

\section{Outlook}

The connection with Meixner-Pollaczek polynomials can be extended to carry
out more detailed computations of Wilson loops in mass-deformed\ ABJ(M)
theory, in the weak coupling limit (some technical aspects are discussed in
more detail in the Appendix). As another open problem, recall that we have
also seen that some of the quotients of partition functions studied above,
such as (\ref{quotient}), which we have related to the Penner matrix model
and hence, in principle, with topological strings, are precisely the ones
relevant in the higher-spin double scaling limit \cite{Hirano:2015yha}. In
this limit $U(N)_{k}\times U(N+M)_{-k}$ ABJ theory with finite $N$ and large 
$M$ and $k$ is conjectured to be dual to $\mathcal{N}=6$ parity-violating
Vasiliev higher spin theory on $AdS_{4}$ with $U(N)$ gauge symmetry \cite%
{Chang:2012kt,Hirano:2015yha}. Recall that, in the oscillator interpretation
put forward here, the $M$ parameter is specifically the Bargmann index $k$
of the positive discrete series representation of $su(1,1)$. It would be
also interesting, therefore, to use the exact solvability found in this
paper to further study this double scaling limit.

Another possible open problem would be to give a physical interpretation of
the appearance of $su(1,1)$ oscillators, for example in terms of motion in $%
AdS$ space\footnote{%
Thanks to Jorge Russo for pointing this out.}. In this sense, it is already
known that some slightly different $su(1,1)$ quantum oscillators emerge as
solutions of Klein-Gordon equation in\ $AdS$ space \cite{Navarro:1996un}.

Finally, and more generally, it would be interesting if a deformation of the 
$su(1,1)$ oscillator eigenfunctions (or equivalently, of the
Meixner-Pollaczek polynomials) can be found, such that the full theory,
given by the matrix model (\ref{tanhmodel}). This deformation would be
seemingly different from the usual $q$-deformation (which holds in the
GUE/Stieltjes-Wigert ensemble description of pure Chern-Simons theory \cite%
{Tierz:2002jj}) because the Vandermonde determinant is deformed according to 
$x\rightarrow \tanh x$, instead of $x\rightarrow \sinh x$. Notice that
different generalizations of the eigenfunctions (beyond the $q$-deformation)
already exist \cite{Jafarov:2012qq}.

\subsection*{Acknowledgements}

Thanks to Jorge Russo and Francesco Aprile for reading the paper and making
valuable comments. This work was supported by the Funda\c{c}\~{a}o para a Ci%
\^{e}ncia e Tecnologia (FCT) through its program Investigador FCT IF2014,
under contract IF/01767/2014.

\newpage

\appendix

\section{Fourier integral and consistency check}

The following Fourier integral \cite{MCS}%
\begin{align}
& \int_{-\infty }^{\infty }e^{-2ixt}P_{n}^{\lambda }\left( x,\phi \right) \
P_{m}^{\lambda }\left( x,\phi \right) \ e^{\left( 2\phi -\pi \right)
x}\left\vert \Gamma \left( \lambda +ix\right) \right\vert ^{2}dx=\frac{%
\Gamma \left( 2\lambda +n\right) \Gamma \left( 2\lambda +m\right) }{%
4^{\lambda }\Gamma \left( 2\lambda \right) n!m!}  \label{Fourier} \\
& \times \frac{2\pi e^{i\pi \lambda }\left( \sinh t\right) ^{n+m}}{\left(
\cos \phi \sinh t+i\sin \phi \cosh t\right) ^{n+m+2\lambda }}\
_{2}F_{1}\left( 
\begin{array}{c}
-n,\ -m \\ 
2\lambda 
\end{array}%
;\ -\left( \frac{\sin \phi }{\sinh t}\right) ^{2}\right) ,  \notag
\end{align}%
is an extension of the orthogonality relationship for Meixner-Pollaczek
polynomials. It suggests that the analytical continuation, required by the
presence of the term in the mass-deformed ABJM theory, namely $e^{-im_{2}x}$%
, can be accounted for with no problems, since (\ref{Fourier}) contains both
a real exponential factor and a Fourier kernel. However, we can explicitly
check that this is the case by explicitly computing the mass-deformed free
energy using an equivalent formulation of the matrix model in terms of a
Hankel determinant and only using the Fourier transform identity (\ref{fu})
to compute the matrix elements. Indeed, since the integral in (\ref{mABJM})
is a determinant%
\begin{equation*}
\frac{1}{N!}\int \prod_{i=1}^{N}\frac{e^{-im_{2}x_{i}}dx_{i}}{\cosh \left(
\pi x_{i}\right) }\prod_{i<j}(x_{i}-x_{j})^{2}=\det \left( c_{i+j}\right)
_{i,j=0}^{N-1}
\end{equation*}%
with matrix elements%
\begin{equation}
c_{n}=\int dx\frac{e^{-im_{2}x}x^{n}}{\cosh \left( \pi x\right) },
\label{Fker}
\end{equation}%
using (\ref{Fourier}) and differentiation under the integral sign (w.r.t $%
m_{2}$), every matrix element is computed explicitly. For $N=2$ for example%
\begin{eqnarray}
\frac{1}{2}\int \prod_{i=1}^{2}\frac{e^{-im_{2}x_{i}}dx_{i}}{\cosh \left(
\pi x_{i}\right) }\prod_{i<j}(x_{i}-x_{j})^{2} &=&\det \left( 
\begin{array}{cc}
1/\cosh (m_{2}/2) & \frac{\sinh (m_{2}/2)}{2i\cosh ^{2}(m_{2}/2)} \\ 
\frac{\sinh (m_{2}/2)}{2i\cosh ^{2}(m_{2}/2)} & \frac{1}{4}\left( \frac{1}{%
\cosh ^{3}(m_{2}/2)}-\frac{\sinh ^{2}(m_{2}/2)}{\cosh ^{3}(m_{2}/2)}\right) 
\end{array}%
\right)   \notag \\
&=&\frac{1}{4\cosh ^{4}(m_{2}/2)}.  \label{detid}
\end{eqnarray}%
The cross-diagonal terms nicely cancel with part of the diagonal term and
the result coincides with the orthogonal polynomial computation (notice that
in this case, Barnes function is $G^{2}(3)=1$). It is also immediate to
check that this determinant is equivalent to carrying out the integrations
on the l.h.s. of (\ref{detid}) explicitly. Let us explicitly also check the $%
N=3$ case%
\begin{equation*}
\det 
\begin{pmatrix}
c_{0} & c_{1} & c_{2} \\ 
c_{1} & c_{2} & c_{3} \\ 
c_{2} & c_{3} & c_{4}%
\end{pmatrix}%
=\frac{1}{16}\func{sech}^{9}\left( m_{2}/2\right) 
\end{equation*}%
where the entries are as in (\ref{detid}) and the additional matrix entries
are%
\begin{eqnarray*}
c_{3} &=&\frac{i}{8}\left( -5\func{sech}^{3}\left( m_{2}/2\right) \tanh
(m_{2}/2\right) +\func{sech}\left( m_{2}/2\right) \tanh ^{3}(m_{2}/2)). \\
c_{4} &=&\frac{5}{16}\func{sech}^{5}\left( m_{2}/2\right) -\frac{9}{8}\func{%
sech}^{3}\left( m_{2}/2\right) \tanh ^{2}(m_{2}/2)+\frac{1}{16}\func{sech}%
\left( m_{2}/2\right) \tanh ^{4}(m_{2}/2).
\end{eqnarray*}%
Thus the same cancellation occurs and only the direct product of the leading
term in the diagonals ends up contributing. This is remarkable but expected
from the theory of orthogonal polynomials and we emphasize that the test was
just to make sure that the Fourier kernel in (\ref{Fker}) (and in the matrix
model) did not invalidate the orthogonal polynomial computation. Notice that
the moments (\ref{Fker}) are all obtained from differentiation, under the
integral sign, of $a_{0}$. This is related to the fact that this Hankel
determinant satisfies a Toda lattice equation, which is a well-known result
in the theory of the six vertex model \cite{BleLiech}, a problem where the
Meixner-Pollaczek polynomials are well-known to provide analytical solutions 
\cite{Colomo:2004jw}.

Finally, note that the argument put forward above, right below Eq. (\ref%
{Fourier}), extends also to the analytical calculability of the Wilson loop
with leads to an average with both types of factors present. Effectively,
the identity (\ref{Fourier}) is analogous to the existing expression for
Hermite polynomials which leads to the celebrated expression for the $\frac{1%
}{2}$-BPS Wilson loop in $\mathcal{N}=4$ in terms of a Laguerre polynomial 
\cite{Drukker:2000rr}. However, as we have pointed out above, the same
result can be obtained (and was indeed obtained decades ago) by using the
action of the creation and annihilation operators on the Fock states. It
would be interesting to do the same in our setting, since the action on the $%
su(1,1)$ Fock basis is equally well-known \cite{Jafarov:2012qq}.

Regarding the use of (\ref{Fourier}) in the evaluation of the Wilson loop
average which, in the mass-deformed case, contains both a factor of the type 
$e^{-2ixt}$ and another one of the type $e^{\left( 2\phi -\pi \right) x}$,
notice that we chose to identify the former with the mass-deformed term $%
e^{-im_{2}x}$. This lead to the expression (\ref{W}). It would be
interesting to study in further detail the convenience of this choice and
also to see if there exists a summation analogous to the one that holds for
the $\frac{1}{2}$-BPS Wilson loop in $\mathcal{N}=4$ SYM, which leads to the
final result in \cite{Drukker:2000rr} and also has lead us to the
interpretation of the Wilson loop in terms of the overlap of two consecutive
displaced number states of the harmonic oscillator (\ref{newW-D}).


\end{document}